\documentclass[12pt]{article}
\usepackage{epsfig,amssymb,euscript,bbold}
\usepackage{amsmath}
\begin{document}

\def\wgnc{\bar{\wedge}}
\def\del{\partial}
\def\der{\overline \del}
\def\wg{\wedge}
\def\gap#1{\vspace{#1 ex}}
\def\tgap{\vspace{3ex}}
\def\sgap{\vspace{5ex}}
\def\lgap{\vspace{20ex}}
\def\half{\frac{1}{2}}
\def\pto{\vfill\eject}
\def\gst{g_{\rm st}}
\def\tC{{\widetilde C}}
\def\x{{\bar x}}
\def\n{\mathtt{n}}
\def\p{p^{\bar x}}
\def\I{{\mathbf I}}
\def\K{{\mathbf K}}
\def\o{{\cal O}}
\def\J{{\cal J}}
\def\S{{\cal S}}
\def\X{{\cal X}}
\def\N{{\cal N}}
\def\M{{\cal M}}
\def\A{{\cal A}}
\def\H{{\cal H}}
\def\F{{\cal F}}
\def\V{\mathtt{V}}
\def\psid{\psi^\dagger}
\def\phid{\phi^\dagger}
\def\ad{a^\dagger}
\def\ddelta{\mathtt{\delta}}
\def\eps{{\epsilon}}
\def\pht{{\tilde \phi}}
\def\D{D}
\def\U{{\cal U}}
\def\V{{\cal V}}
\def\W{{\cal W}}
\def\u{u}
\def\d{{\cal D}}
\def\myitem#1{\gap2\noindent\underbar{#1}\gap2}
\def\winf{\ensuremath{W_\infty}}
\def\ads{AdS$_5 \times$S$^5$}


\def\be{\begin{equation}}
\def\ee{\end{equation}}
\def\ba{\begin{array}{l}}
\def\ea{\end{array}}
\def\bea{\begin{eqnarray}}
\def\eea{\end{eqnarray}}
\def\beas{\begin{eqnarray*}}
\def\eeas{\end{eqnarray*}}
\def\eq#1{(\ref{#1})}
\def\fig#1{Fig \ref{#1}} 
\def\re#1{{\bf #1}}
\def\bull{$\bullet$}
\def\ub{\underbar}
\def\nn{\nonumber\\}
\def\nl{\hfill\break}
\def\ni{\noindent}
\def\bibi{\bibitem}
\def\ket#1{| #1 \rangle}
\def\bra#1{ \langle #1 |}
\def\overlap#1#2{\langle #1 | #2 \rangle}
\def\mat#1#2#3{ \langle #1 | #2 | #3 \rangle}
\def\vev#1{\langle #1 \rangle} 
\def\lsim{\stackrel{<}{\sim}}
\def\gsim{\stackrel{>}{\sim}}
\def\mattwo#1#2#3#4{\left(\begin{array}{cc}#1&#2\\#3&#4\end{array}\right)} 
\def\tgen#1{T^{#1}}

\def\gap#1{\vspace{#1 ex}}
\def\tgap{\vspace{3ex}}
\def\sgap{\vspace{5ex}}
\def\lgap{\vspace{20ex}}
\def\half{\frac{1}{2}}
\def\pto{\vfill\eject}
\def\psid{\psi^\dagger}
\def\phid{\phi^\dagger}
\def\ad{a^\dagger}
\def\sigmad{\sigma^\dagger}
\def\betad{\beta^\dagger}
\def\ddelta{\mathtt{\delta}}
\def\eps{{\epsilon}}
\def\dcff#1#2{D(#1| #2)}
\def\gcff#1#2{\Gamma(#1| #2)}
\def\tgcff#1#2{\tilde \Gamma(#1| #2)}
\def\ads{$AdS_5 \times S^5$}

\title{
{\baselineskip -.2in
\vbox{\small\hskip 4in \hbox{hep-th/0606088}}
\vbox{\small\hskip 4in \hbox{TIFR/TH/06-14}}
\vbox{\small\hskip 4in \hbox{}}}
\vskip .4in
Counting 1/8-BPS Dual-Giants}
\author{Gautam Mandal $^1$ and Nemani V. Suryanarayana $^2$\\
{}\\
{\small{\it $^1$ Department of Theoretical
    Physics,}}\\
{\small{\it TIFR, Homi Bhabha Road, Mumbai, 400005, India.}} \\
{\small{E-mail: {\tt mandal@theory.tifr.res.in}}} \\
{\small{$^2$ {\it Perimeter Institute for Theoretical
      Physics,}}}\\
{\small{\it 31 Caroline Street North, Waterloo, ON, N2L 2Y5,
    Canada}}\\
{\small{E-mail: {\tt vnemani@perimeterinstitute.ca}}}
}
\maketitle

\abstract{We count 1/8-BPS states in type IIB string
  theory on $AdS_5 \times S^5$ background which carry three
  independent angular momenta on $S^5$. These states can be counted by
  considering configurations of multiple dual-giant gravitons up to
  $N$ in number which share at least four supersymmetries. We map this
  counting problem to that of counting the energy eigenstates
  of a system of $N$ bosons in a 3-dimensional harmonic oscillator. We
  also count 1/8-BPS states with two independent non-zero spins in
  $AdS_5$ and one non-zero angular momentum on $S^5$ by considering
  configurations of arbitrary number of giant gravitons that share at
  least four supersymmetries.}

\newpage

\tableofcontents

\section{Introduction}

In the context of AdS/CFT correspondence \cite{jmm} it is of interest
to count the number of BPS states for fixed charges and
supersymmetries in both ${\cal N} = 4$ $U(N)$ SYM and in type IIB
string theory on $AdS_5 \times S^5$ background with applications to
black holes in mind. The most interesting outstanding problem in this
context is to count the microstates of the supersymmetric black holes
in $AdS_5$. These black holes were first found by Gutowski and Reall
\cite{gr} (and were further generalized in \cite{cclp, klr}) and when
lifted to 10 dimensions preserve just 2 supersymmetries
\cite{ggs}. This program of counting BPS states on both sides of the
AdS/CFT correspondence has only been achieved so far for the simplest
case of half-BPS states that carry one $U(1)$ R-charge. A generic
state in string theory on $AdS_5 \times S^5$ can be specified by
giving the quantum numbers $(E, S_1, S_2, J_1, J_2, J_3)$ where $E$ is
the energy in global $AdS_5$, $S_1$ and $S_2$ are the two independent
angular momenta in $AdS_5$ and $(J_1, J_2, J_3)$ are the three
independent R-charges corresponding angular momenta in $S^5$.

In the case of half-BPS states from the string theory point of view
one can count either the multi-giant graviton states or
multi-dual-giant graviton states \cite{nvs}. A giant graviton is a
half-BPS classical D3-brane configurations wrapping an $S^3 \subset
S^5$ and rotating along one of the transverse directions to it within
$S^5$ \cite{mst}. A dual-giant graviton is another half-BPS D3-brane
configuration that wraps an $S^3 \subset AdS_5$ and rotating along a
maximal circle of $S^5$ \cite{gmt, hhi}. From the point of view of
giants the stringy exclusion principle manifests itself in the fact
that the maximum angular momentum that a single giant graviton can
carry is $N$. The same stringy exclusion principle from the point of
view of dual-giant gravitons is that there is an upper limit on the
number of dual-giants which is again given by $N$. From the
perspective of supergravity and probe branes this upper limit on the
number of dual-giants has to do with the way the 5-form RR-flux
decreases inside each dual-giant by one unit (see for instance
\cite{bs, nvs}). To count the half-BPS states one can consider an
arbitrary number of half-BPS giants, treated as bosons, in an
$N$-level equally spaced spectrum. The same half-BPS state counting
can be done by counting configurations of multiple dual-giant
gravitons, treated as bosons, in an infinite equally spaced spectrum
with an upper limit on the dual-giants given by $N$.

Progress in counting BPS states with lower supersymmetry in the SYM
has been made difficult by the fact that the number of these states is
known to jump from zero YM coupling to non-zero coupling. In this note
we aim to count a subclass of 1/8-BPS states in type IIB string theory
on $AdS_5 \times S^5$ geometry that carry three independent $U(1)$
R-charges $J_1$, $J_2$ and $J_3$ and have the energy (conjugate to the
global $AdS_5$ time) $E = J_1 + J_2 + J_3$. This subclass of 1/8-BPS
states preserve the $SO(4)$ invariance of the $S^3 \subset AdS_5$. We
will do the counting of 1/8-BPS states by considering all possible
multiple dual-giant graviton configurations which preserve a common
set of at least 4 supersymmetries. Since we are interested in putting
together half-BPS dual-giants to obtain the 1/8-BPS states the
argument that there should be an upper limit on the number of
dual-giants given by $N$ continues to be valid.

In \cite{mikhailov} Mikhailov described the most general 1/8-BPS giant
gravitons as the intersections of holomorphic complex surfaces in
${\mathbb C}^3$ with $S^5$. So for counting the 1/8-BPS states one may
try to quantize these giant graviton configurations and count them. We
have been informed that this has been recently achieved by Biswas,
Gaiotto, Lahiri and Minwalla \cite{bglm}. Our results agree with
theirs. The agreement for 1/8-BPS states suggests that both
Mikhailov's giants and our dual-giants present dual descriptions of
the same set of states like in the half-BPS case.

We also consider 1/8-BPS states which carry non-zero $(S_1, S_2, J_1)$
and have $E = S_1 + S_2 + J_1$. This time we consider multiple
giant-graviton configurations which have angular momenta in $AdS_5$
part as well that share at least 4 supersymmetries.

The rest of this note is organized as follows. In section 2 we
construct a class of dual-giant solutions which share at least 4
supersymmetries. We show that this class of solutions is the full set
of 1/2-BPS dual-giants that share a given 4 supersymmetries with the
rest in their class. We find the solution space and a sympletic form
on this space and quantize it. In section 3 we give the partition
function of the 1/8-BPS states with $(J_1, J_2, J_3)$ charges. In
section 4 we consider the problem of counting 1/8-BPS states with
$(S_1, S_2, J_1)$ charges. In section 5 we conclude with a summary and
some remarks.

\section{1/8-BPS dual-giant configurations}

In this section we will find the most general dual-giant
configurations which preserve a given 4 supersymmetries in $AdS_5
\times S^5$. As mentioned in the introduction we put half-BPS
dual-giants together to make the lower supersymmetric
configurations. A given half-BPS dual-giant has momentum along one of
the maximal circles of $S^5$. A general dual-giant configuration
contains dual-giants rotating along different maximal circles. There
is no a priori guarantee that such a generic configuration preserves
any supersymmetries. Below we will look for a class of single
dual-giant solutions which shares at least 4 supersymmetries with the
first one.

\subsection{The dual-giant solutions}

We regard $S^5$ as a submanifold in ${\mathbb C}^3$ with coordinates
\begin{equation}
\label{c-3}
(z_1= l \mu_1 \, e^{i \, \xi_1}, z_2 = l \mu_2 \, e^{i \, \xi_2}, z_3
= l \, \mu_3 \, e^{i \, \xi_3})
\end{equation}
as $z_i \bar z_i = l^2$ and $(\mu_1,
\mu_2, \mu_3)
= (\sin\alpha, \cos\alpha \sin\beta, \cos\alpha \cos\beta)$:
\bea
ds^2|_{S^5} = l^2\left( d\alpha^2 + \cos^2\alpha \beta^2 +
\mu_i^2 d\xi_i^2 \right)
=  dz_i d{\bar z}_i
\label{s-5}
\eea
The vielbeins are
\begin{eqnarray}
\label{vbsone}
&&e^0= V^{1/2}(r) \, dt, ~~ e^1 = V^{-1/2}(r) \, dr, ~~ e^2 = r\,
d\theta,\cr
&&e^3 = r\, \nu_1 \, d\phi_1, ~~ e^4 = r\, \nu_2\, d\phi_2, ~~ e^5 = l\,
d\alpha, ~~ e^6 = l \, \cos\alpha \, d\beta, \cr
&&e^7 = l\, \mu_1\, d\xi_1, ~~ e^8 = l\, \mu_2 \, d\xi_2, ~~ e^9 = l\,
\mu_3 \, d\xi_3.
\end{eqnarray}
where $\nu_1 = \cos\theta$, $\nu_2 = \sin\theta, V(r) = 1 + r^2/l^2$.
The five form RR field strength is
\begin{equation}
\label{fiveform}
F^{(5)} = - \frac{4}{l} \left[ e^0 \wedge e^1 \wedge e^2 \wedge e^3
\wedge e^4 + e^5 \wedge e^6 \wedge e^7 \wedge e^8 \wedge e^9 \right].
\end{equation}
The 4-form potential can be written as
\begin{equation}
C^{(4)} = r^4 \, \cos\theta \, \sin\theta \, d(t/l) \wedge d\theta \wedge
d\phi_1 \wedge d\phi_2 + l^4 \, \cos^4 \alpha \, \sin\beta \,
\cos\beta \, d\beta
\wedge d\xi_1 \wedge d\xi_2 \wedge d\xi_3.
\end{equation}
The general embedding of a D3-brane wrapped on the $S^3 \in
AdS_5$ is given by 
\begin{eqnarray}
&&  t = \sigma_0=\tau,\,  r=r(\tau), \, \theta = \sigma_1, \, \phi_1 =
  \sigma_2, \, \phi_2 = \sigma_3, 
\cr
&&\alpha= \alpha(\tau), \, \beta= \beta(\tau), \,  
\xi_1 = \xi_1(\tau), \, \xi_2 = \xi_2(\tau), \, \xi_3 = \xi_3(\tau). 
\end{eqnarray}
The pull-back of $C^{(4)}$ onto the world-volume
is $C^{(4)}_{\sigma_0 \sigma_1 \sigma_2 \sigma_3} = l^{-1} r^4 \,
\cos\theta \, \sin\theta$. The DBI and WZ lagrangian density is
\begin{equation}
{\cal L} = -\frac{N}{2 \, \pi^2 \, l^4} r^3 \cos\sigma_1 \sin\sigma_1
\left[\Delta^{1/2} - \frac{r}{l} \right]
\end{equation}
where 
\be
\Delta = V(r) - \frac{{\dot r}^2}{ V(r)} - l^2 (
{\dot \alpha}^2 + \cos^2\alpha \, {\dot \beta}^2 + \mu_1^2 \, \dot
\xi_1^2 + \mu_2^2 \, \dot \xi_2^2 + \mu_3^2 \, \dot \xi_3^2).
\ee 
We can integrate over the angles $\sigma_i$ to get Action $= \int
dt\, L$ where the effective point-particle Lagrangian $L$ is 
\begin{eqnarray}
\label{d3lag}
 L = \int d\sigma_1 \, d\sigma_2 \, d\sigma_3 ~\, {\cal L} =
-\frac{N}{l^4} r^3 \left[ \Delta^{1/2} - \frac{r}{l} \right]
\end{eqnarray}
The conjugate variables of the classical mechanics model are
\begin{eqnarray}
\label{momenta}
P_r &&= \frac{N \, r^3 \, \dot r}{l^4 \, V(r) \, \Delta^{1/2}},~~
P_\alpha = \frac{N \, r^3 \, \dot \alpha}{l^2 \, \Delta^{1/2}},~~
P_\beta = \frac{N \, r^3 \, \cos^2\alpha \, \dot \beta}{l^2 \,
  \Delta^{1/2}}, \cr
P_{\xi_1} &&= \frac{N \, r^3 \, \mu_1^2 \, \dot \xi_1}{l^2
  \Delta^{1/2}}, ~~ 
P_{\xi_2} = \frac{N \, r^3 \, \mu_2^2 \, \dot \xi_2}{l^2 \Delta^{1/2}},
P_{\xi_3} = \frac{N \, r^3 \, \mu_3^2 \, \dot \xi_3}{l^2 \Delta^{1/2}}
\end{eqnarray}
By looking at the equations of motion of the lagrangian in
(\ref{d3lag}) it is easy to see that setting
\begin{equation}
\label{dgsolns}
\dot r = \dot \alpha = \dot \beta = 0, ~~ |\dot \xi_1| = |\dot \xi_2|
= | \dot \xi_3| = 1/l
\end{equation}
solves them.

\subsection{Kappa projections}
In this subsection we will show that the solutions in (\ref{dgsolns})
are all supersymmetric. We will see that depending on the signs of
$\dot \xi_i$ there are 8 disjoint sets of these supersymmetric
dual-giant solutions which do not share any common
supersymmetries. Different dual-giant solutions corresponding to
different values of $r$, $\alpha$, $\beta$ and $\xi_i (\tau=0)$ but
with fixed signs for $\dot \xi_i$ preserve at least 4 common
supersymmetries of the background $AdS_5 \times S^5$ geometry.

To write down the kappa projection equation for the probe D3-brane we
need the world-volume gamma matrices, which are
\begin{eqnarray}
\label{wvgammas}
\gamma_{\tau} &=& V^{1/2}(r) \, \Gamma_0
+ \frac{\dot r }{V^{1/2}(r)} \, \Gamma_1 + l \, (\dot \alpha \,
\Gamma_5 + \cos\alpha \, \dot \beta \, \Gamma_6  
+ \sum_{i=1}^3 \dot \xi_i \, \mu_i \, \Gamma_{6+i})
\cr
\gamma_{\sigma_1} &=& r \, \Gamma_2, ~~~ \gamma_{\sigma_2} = r \,
\cos\sigma_1 \, \Gamma_3, ~~~ \gamma_{\sigma_3} = r \, \sin\sigma_1 
\, \Gamma_4
\end{eqnarray}
where $\Gamma_a$ are the 10-dimensional tangent space gamma matrices
satisfying $\{\Gamma_a, \Gamma_b \} = 2 \, \eta_{ab}$. The
world-volume gamma matrices of (\ref{wvgammas}) satisfy $\{ \gamma_m,
\gamma_n \} = 2\, g_{mn}$. The kappa projection matrix is:
\begin{eqnarray}
\label{kappamatrix}
&& \! \! \! \! \! \! \! \! \! \Gamma = \frac{1}{4! 
\sqrt{-\det\, g_{mn}}}\, 
\epsilon^{mnpq} \gamma_{mnpq} \cr
&& \! \! \! \! \! \!  = \Delta^{-1/2} \left[ V^{1/2}(r) \, \Gamma_0 + 
\frac{\dot r}{ V^{1/2}(r)} \, \Gamma_1 +  l\, (\dot \alpha \Gamma_5 +  
\cos\alpha \, \dot \beta \, \Gamma_6 + \sum_{i=1}^3 \dot\xi_i \, \mu_i 
\, \Gamma_{6+i}) \right] \Gamma_{234}
\nn
\end{eqnarray}
where $\Delta$ is defined as before. With this the kappa projection on
the Killing spinor $\epsilon$ of the background $AdS_5 \times S^5$
geometry is
\begin{equation}
\label{kapro}
\Gamma \, \epsilon = i \, \epsilon.
\end{equation}
The chirality convention for $\epsilon$ is
\begin{equation}
\label{chirality}
\Gamma_0 \cdots \Gamma_9 \epsilon = - \epsilon.
\end{equation}
The Killing spinor equations of $AdS_5 \times S^5$ are:
\begin{equation}
\label{Kseqs}
D_\mu \epsilon + \frac{i}{1920} F^{(5)}_{\nu_1 \nu_2 \nu_3 \nu_4
\nu_5} \Gamma^{\nu_1 \nu_2 \nu_3 \nu_4 \nu_5} \Gamma_\mu \, \epsilon
= 0
\end{equation}
%
%
%
The solution of these equations can be written as:
\begin{eqnarray}
\label{adskss}
\epsilon &=& ~ e^{\frac{i}{2} \alpha \, \Gamma_5 \, \tilde \gamma} \,
e^{\frac{i}{2} \beta \, \Gamma_6 \tilde \gamma} \, e^{\frac{1}{2}
\xi_1 \Gamma_{57}} \, e^{\frac{1}{2} \xi_2  \Gamma_{68}} \,
e^{\frac{i}{2} \xi_3  \Gamma_9 \, \tilde \gamma} \cr
&& \times e^{\frac{i}{2} \sinh^{-1}\! (\frac{r}{l}) \, \Gamma_1 \,
\gamma} \, e^{\frac{i}{2l} t \, \Gamma_0 \, \gamma} \, e^{\frac{1}{2}
\theta \, \Gamma_{12}} \, e^{\frac{1}{2} \phi_1 \Gamma_{13}} \,
e^{\frac{1}{2} \phi_2 \Gamma_{24}} \, \epsilon_0 \equiv M \,
\epsilon_0.
\end{eqnarray}
where $\gamma = \Gamma^{01234}$ and $\tilde \gamma =
\Gamma^{56789}$. The full kappa projection equation is then reads
\begin{equation}
\label{fullkpeqn}
\left[ V^{1/2}(r) \, \Gamma_0 
-i \, \Delta^{1/2} \, \Gamma_{234}
+ \frac{\dot r}{V^{1/2}(r)} \, \Gamma_1 + l \, (\dot \alpha
\,\Gamma_5 + \cos\alpha \,\dot \beta \, \Gamma_6 + 
\sum_{i=1}^3 \dot \xi_i \, \mu_i \, \Gamma_{6+i} ) \right] M
\epsilon_0 = 0 
\end{equation}
We now show that the solutions in (\ref{dgsolns}) are
supersymmetric. Let us first choose the signs of $\xi_i$'s to be the
same and positive. On the solution the kappa projection equation
becomes
\begin{equation}
\label{kponsolns}
\left[ V^{1/2}(r_0) \, \Gamma_0 -i \frac{r_0}{l} \Gamma_{234}
+ \mu_1 \, \Gamma_7 + \mu_2 \, \Gamma_8  + \mu_3 \, \Gamma_9
\right] M \epsilon_0 = 0
\end{equation}
where $r_0$ is the value of $r$ in the solutions. Some useful
intermediate steps in this calculation in simplifying this equation
are:
\begin{eqnarray}
\label{intsteps}
&&\Gamma_0 M = M \left[ V^{1/2} \Gamma_0 - i \frac{r_0}{l} e^{-i
\frac{t}{l} \Gamma_0 \gamma} \left( \cos\theta \, e^{- \phi_1
\Gamma_{13}} \Gamma_1 + \sin\theta \, e^{- \phi_2 \Gamma_{24}}
\Gamma_2 \right) \Gamma_0 \, \gamma \right], \cr
&&\Gamma_2 M = M \left[ V^{1/2} e^{-i \frac{t}{l} \Gamma_0 \gamma}
\left( \cos\theta \, e^{-\phi_2 \Gamma_{24}} \Gamma_2
- \sin\theta \, e^{-\phi_1 \Gamma_{13}} \Gamma_1 \right) -i
  \frac{r}{l} e^{-\phi_1 \Gamma_{13} - \phi_2 \Gamma_{24}} \Gamma_{12}
  \gamma \right], \cr
&&\Gamma_3 M = M \left[ V^{1/2} \, e^{-\phi_1 \Gamma_{13}} \, e^{-i
\frac{t}{l} \Gamma_0 \gamma} \Gamma_3 - i \frac{r_0}{l} \left(
\cos\theta \, \Gamma_1 + \sin\theta \, e^{- (\phi_1 \Gamma_{13} +
\phi_2 \Gamma_{24})} \Gamma_2 \right) \Gamma_3 \gamma \right], \cr
&&\Gamma_4 M = M \left[ V^{1/2}\, e^{- \phi_2 \Gamma_{24}} \, e^{-
i \frac{t}{l} \Gamma_0 \gamma} \Gamma_4 - i \frac{r_0}{l} \left(
\cos\theta \, e^{- (\phi_1 \Gamma_{13}+ \phi_2 \Gamma_{24})} \Gamma_1
+ \sin\theta \, \Gamma_2 \right) \Gamma_4 \gamma \right], \cr
&&\Gamma_7 M = \cr
&& M \Big[ \cos\alpha \, \cos\beta \, e^{-\xi_1
\Gamma_{57}} \, e^{-i \xi_3 \Gamma_9 \tilde \gamma} \Gamma_7
- i \cos\alpha \, \sin\beta \, e^{-\xi_1 \Gamma_{57}} e^{-\xi_2
\Gamma_{68}} \Gamma_{67} \tilde \gamma - i \sin\alpha \, \Gamma_{57}
\tilde \gamma \Big], \cr
&&\Gamma_8 M = \cr
&& M \Big[ \cos\alpha \, \cos\beta \, e^{-i \xi_3
\Gamma_9 \tilde \gamma} e^{- \xi_2 \Gamma_{68}} \Gamma_8 - i
\cos\alpha \, \sin \beta \, \Gamma_{68} \tilde \gamma - i \sin \alpha
\, e^{- \xi_2 \Gamma_{68}} e^{- \xi_1
\Gamma_{57}} \Gamma_{58} \tilde \gamma \Big],\nonumber
\cr
&&\Gamma_9 M = \cr
&& M \Big[ \cos\alpha \, \cos \beta \, \Gamma_9 - i
\cos\alpha \, \sin\beta \, e^{-i \xi_3 \Gamma_9 \tilde \gamma} e^{-
\xi_2 \Gamma_{68}} \Gamma_{69} \tilde \gamma - i \sin\alpha \, e^{-i
\xi_3 \Gamma_9 \tilde \gamma} e^{- \xi_1 \Gamma_{57}} \Gamma_{59} \tilde
\gamma \Big].
\end{eqnarray}
Let us also register the following identity:
\begin{eqnarray}
\Gamma_{234}M &=& M \left[ V^{1/2} e^{-i \frac{t}{l} \Gamma_0 \gamma}
\left( \cos\theta \, e^{-\phi_1 \Gamma_{13}} \Gamma_{234} - \sin\theta
e^{-\phi_2 \Gamma_{24}} \Gamma_{134} \right) - i \frac{r}{l}
\Gamma_{1234} \gamma \right] \cr
&=& - M \left[ V^{1/2} e^{-i \frac{t}{l} \Gamma_0 \gamma} \left(
\cos\theta \, e^{-\phi_1 \Gamma_{13}} \Gamma_1 + \sin\theta
e^{-\phi_2 \Gamma_{24}} \Gamma_2 \right) \Gamma_0 \gamma + i \frac{r}{l}
\Gamma_0 \right]
\end{eqnarray}
so that we have
\begin{equation}
\left[ V^{1/2} \Gamma_0 - i \frac{r}{l} \Gamma_{234} \right] \epsilon =
M \Gamma_0 \epsilon_0.
\end{equation}
Using these identities eq.(\ref{kponsolns}) can be rewritten as
\begin{eqnarray}
\label{preproj}
&& M \Big[\Gamma_0 -i \mu_1 \, \mu_2 \, e^{- \xi_1
\Gamma_{57} - \xi_2 \Gamma_{68}} \tilde \gamma (\Gamma_{67} +
\Gamma_{58}) 
+ \mu_1 \, \mu_3 \, e^{-\xi_1 \Gamma_{57}} \, e^{-i \xi_3
  \Gamma_9 \tilde \gamma} (\Gamma_7 - i \Gamma_{59} \tilde \gamma) \cr
&&
+ \mu_2 \, \mu_3 \, e^{-i \xi_3 \Gamma_9 \tilde \gamma} \, e^{-\xi_2 \,
  \Gamma_{68}} \, ( \Gamma_8 - i \, \Gamma_{69} \, \tilde \gamma) - i
\tilde \gamma \, (\mu_1^2 \, \Gamma_{57} + \mu_2^2 \, \Gamma_{68}) +
\mu_3^2 \, \Gamma_9 \Big] \epsilon_0 = 0.
\end{eqnarray}
This equation can be solved by imposing the following projections on
$\epsilon_0$
\begin{eqnarray}
&& (\Gamma_{67} + \Gamma_{58}) \epsilon_0 = 0,  
~~ (\Gamma_7 - i \,\Gamma_{59} \, \tilde \gamma) \epsilon_0 = 0, 
~~ (\Gamma_8 - i \, \Gamma_{69} \, \tilde \gamma) \epsilon_0 = 0, \cr
&& (\Gamma_0 -i
  \Gamma_{57} \tilde 
\gamma) \epsilon_0 = 0,~~ (\Gamma_9 + i \, \Gamma_{57} \, \tilde
  \gamma) \epsilon_0 =0.
\end{eqnarray}
It is easy to see that out of these the independent projections are
(as the second and the fourth projections are equivalent and so are
third and fifth)
\begin{equation}
(\Gamma_0 - i \Gamma_{57} \tilde \gamma) \epsilon_0 = 0, ~~~
(\Gamma_0 - i \Gamma_{68} \tilde \gamma) \epsilon_0 = 0, ~~~
(\Gamma_0 + \Gamma_9) \epsilon_0 = 0.
\label{susy-projections}
\end{equation}
By considering the special cases of setting two of $\dot \xi_i$ to
zero and finding the kappa projections one recognizes these three
projection conditions as the ones for half-BPS dual-giants which
rotate in $z_1$, $z_2$, $z_3$ planes respectively. One can further see
that different signs for $\dot \xi_i$ result in similar equations with
different relative signs in eqs.(\ref{susy-projections}) corresponding
to reversing the direction of motion of some of the three half-BPS
dual giants in their respective planes. That is, for the solution with
\be
(l \, \dot \xi_1, l \, \dot \xi_2, l \, \dot \xi_3) = (\lambda_1,
\lambda_2, \lambda_3), \quad \lambda_i = \pm 1
\label{signs}
\ee
we have
\begin{equation}
(\Gamma_0 - i \,\lambda_1 \, \Gamma_{57} \tilde \gamma) \epsilon_0 =
  0, ~~~ 
(\Gamma_0 - i \,\lambda_2 \,\Gamma_{68} \tilde \gamma) \epsilon_0 = 0,
  ~~~ 
(\Gamma_0 + \,\lambda_3 \,\Gamma_9) \epsilon_0 = 0.
\label{kappa-lambda}
\end{equation}
Thus we have seen that the solutions in (\ref{dgsolns}) with all signs
of $\dot \xi_i$ being positive preserve at least 4 supersymmetries
consistent with the projections in (\ref{susy-projections}) for any
arbitrary values of $\alpha$, $\beta$ and $r$. However one should note
that for fixed values of $\alpha$, $\beta$ and $r$ the projection
equation (\ref{preproj}) can give 16 supersymmetries. This can be seen by
redefining the tangent space $\Gamma$-matrices appropriately. 

In the next subsection we will prove the converse, namely that these
are \underline{all} the dual-giant solutions which preserve at least
these 4 given supersymmetries.

\subsection{Necessity of the conditions \eq{dgsolns}}

In this subsection, unlike the previous one, we will not assume the
solutions \eq{dgsolns} to begin with. We will rather show that these
solutions (with signs \eq{signs}) follow uniquely if we demand the
kappa projections
\eq{kapro},\eq{kappamatrix} on the 1/8-th supersymmetric
subspace of spinors $\epsilon$ defined by
\eq{adskss},\eq{kappa-lambda}. This will therefore show that
those solutions are the full set of solutions consistent
with the specified set of 4 supersymmetries.

To begin, we rewrite the supersymmetry projections 
\eq{susy-projections} (we will assume all $\lambda_i$'s to
be $+1$, the generalization to arbitrary signs being straightforward)
\begin{equation}
\label{kpeqns2}
(\Gamma_0 - i \Gamma_{57} \tilde \gamma) \, \epsilon_0 = 0, ~~~
(\Gamma_0 - i \Gamma_{68} \tilde \gamma) \, \epsilon_0 = 0, ~~~
(\Gamma_0 + \Gamma_9) \, \epsilon_0= 0.
\end{equation}
We can use these to simplify the full expression (\ref{adskss}) of
the Killing spinor on $AdS_5 \times S^5$ to
\begin{eqnarray}
\epsilon &=&\times e^{\frac{i}{2} \sinh^{-1}\! (\frac{r}{l}) \,
  \Gamma_1 \, \gamma} \, e^{\frac{i}{2}
  (\frac{t}{l} + \xi_1 + \xi_2 + \xi_3) \, \Gamma_0 \, \gamma} \,
  e^{\frac{1}{2} 
\theta \, \Gamma_{12}} \, e^{\frac{1}{2} \phi_1 \Gamma_{13}} \,
e^{\frac{1}{2} \phi_2 \Gamma_{24}} \cr
&& ~ e^{\frac{i}{2} \alpha \, \Gamma_5 \, \tilde \gamma} \,
e^{\frac{i}{2} \beta \, \Gamma_6 \tilde \gamma} 
\, \epsilon_0 \equiv {\hat M} \,
\epsilon_0.
\end{eqnarray}
To simplify the kappa projection equations (\ref{fullkpeqn}) we will
need
\begin{eqnarray}
\Gamma_1 \, {\hat M} &=& {\hat M} \, e^{-i\frac{t}{l} \Gamma_0 \gamma}
      [ \cos\theta \, e^{-\phi_1 \, \Gamma_{13}} \Gamma_1 + \sin\theta
      \, e^{-\phi_2 \, \Gamma_{24}} \, \Gamma_2], \cr
\Gamma_5 \, {\hat M} &=& {\hat M} \, [ \cos\beta \, \Gamma_5 + i
      \,\sin\beta \, \Gamma_{56} \, \tilde \gamma] \cr
\Gamma_6 \, {\hat M} &=& {\hat M} \, [ \cos\alpha \, \Gamma_6 - i \,
      \sin\alpha \, \cos\beta \, \Gamma_{56} \tilde \gamma +
      \sin\alpha \, \sin\beta \, \Gamma_5], \cr
\Gamma_7 \, {\hat M} &=& {\hat M} \, [ \mu_3 \, \Gamma_7 - i \, \mu_2
      \, \Gamma_{67} \, \tilde \gamma - i \, \mu_1 \, \Gamma_{57} \,
      \tilde \gamma]\cr
\Gamma_8 \, {\hat M} &=& {\hat M} \, [ \mu_3 \, \Gamma_8 - i \, \mu_2
      \, \Gamma_{68} \, \tilde \gamma - i \, \mu_1 \, \Gamma_{58} \,
      \tilde \gamma] \cr
\Gamma_9 \, {\hat M} &=& {\hat M} \, [ \mu_3 \, \Gamma_9 - i \, \mu_2
      \, \Gamma_{69} \, \tilde \gamma - i \, \mu_1 \, \Gamma_{59} \,
      \tilde \gamma]
\end{eqnarray}
along with $\Gamma_0 \, M$ and $\Gamma_{234} \, M$ relations from the
earlier section. Since we want to demand the projections in
(\ref{kpeqns2}) to be valid at every point of the world-volume of the
D3-brane and so for convenience we can set $\tau=0$ and $\sigma_1
= \sigma_2 = 0$. Then the first three terms in Eq.(\ref{fullkpeqn})
can be simplified to
\begin{equation}
[V^{1/2}\Gamma_0 + \frac{\dot r}{V^{1/2}} \Gamma_1 - i \Delta^{1/2}
  \Gamma_{234}] \hat M = {\hat M} [ (V-\frac{r}{l} \Delta^{1/2})
  \Gamma_0 + \frac{\dot r}{V^{1/2}} + i \Gamma_{01} \gamma
  V^{1/2}(\frac{r}{l} - \Delta^{1/2})]
\end{equation}
The last five terms in eq.(\ref{fullkpeqn}) can be simplified to 
\begin{eqnarray}
&& \!\!\!\!\!\!\!\!\!\!
[\dot \alpha \, \Gamma_5 + \cos\alpha \, \dot \beta \, \Gamma_6 + 
  \sum_{i=1}^3 \mu_i \, \dot \xi_i \, \Gamma_{6+i}] \, {\hat M} \,
= {\hat M} \, [ (\cos\beta \, \dot \alpha + \sin\alpha \, \cos\alpha
  \, \sin\beta \, \dot \beta) \, \Gamma_5 \cr
&& \!\!\!\!\!\!\!\!\!\! + \Gamma_6 \, \cos^2\alpha
  \, \dot \beta + i \, \Gamma_{56} \, \tilde \gamma \, (\sin\beta \,
  \dot \alpha - \sin\alpha \, \cos\alpha \, \cos\beta \, \dot \beta) +
  i \, \Gamma_{59} \, \tilde \gamma \, \mu_1 \, \mu_2 \, (\dot \xi_1 -
  \dot \xi_3) \cr
&&\!\!\!\!\!\!\!\!\!\!  - i \, \Gamma_{58} \, \tilde \gamma \, \mu_1
  \mu_2 \, (\dot \xi_2 - \dot \xi_1) - i \, \Gamma_{69} \, \tilde
  \gamma \, ( \dot \xi_2 - \dot \xi_3) - i \, \mu_1^2 \, \dot \xi_1 \,
  \Gamma_{57} \tilde \gamma - i \, \mu_2^2 \, \dot \xi_2 \,
  \Gamma_{68} \, \tilde  \gamma + \mu_3^2 \, \dot \xi_3 \, \Gamma_9]
\end{eqnarray}
Thus the full kappa projection equation (\ref{fullkpeqn}) is a linear
combination of various products of 10-dimensional $\gamma$-matrices
for the tangent space acting on a constant spinor $\epsilon_0$ which
is arbitrary but for the projection equations (\ref{kpeqns2}). Recall
that a linearly independent basis of these $32 \times 32$ matrices is
given by $\Gamma_{m \, n \, \cdots}$'s. Since $\epsilon_0$ is
arbitrary only upto the projections in eq.(\ref{kpeqns2}) we have to
use them to eliminate operators which kill $\epsilon_0$ and then put
the coefficients of the remaining independent $\Gamma_{m \, n \,
\cdots}$ to zero. Doing this we see that $\dot \beta = 0$ as the
coefficient of $\Gamma_6$ (which does not appear anywhere else), which
then implies $\dot \alpha = 0$. Next notice that $\Gamma_{58}$,
$\Gamma_{59}$ and $\Gamma_{69}$ would not occur anywhere else and so
we have to set $\dot \xi_1 = \dot \xi_2 = \dot \xi_3 \equiv
\omega/l$. Note that these are not absolute values - unlike in the
analysis of equations of motion.  Since $\dot r$ appears with
$\Gamma_1$ and $\Gamma_1$ does not occur anywhere else $\dot r =
0$. The coefficient of $\Gamma_{01} \gamma$ has to be set to zero too
and this implies $\omega = \pm 1$. Finally using the projection
equations (\ref{kpeqns2}) we can convert $\Gamma_{57} \tilde \gamma$,
$\Gamma_{68} \, \tilde \gamma$ and $\Gamma_9$ into $\Gamma_0$. Using
the relations so far and then setting the coefficient of $\Gamma_0$ to
zero requires $\omega = 1$. That completes the proof that the set of
solutions we found is complete.

\subsection{\label{sec:symplectic} Reduced phase space}

The supersymmetry constraints \eq{kpeqns2} translate to the following
in terms of the canonical momenta (see \eq{momenta})
\begin{eqnarray}
&& P_r = 0, ~~
P_\alpha = 0, ~~
P_\beta = 0,
\nn
&& P_{\xi_1} - \frac{N}{l^2} r^2 \, \mu_1^2=0 , ~~
P_{\xi_2} -  \frac{N}{l^2}   r^2 \, \mu_2^2=0,
P_{\xi_3} - \frac{N}{l^2}   r^2 \, \mu_3^2=0
\label{constraints}
\end{eqnarray}
Thus, the original 12-dimensional phase space is reduced to a
six-dimensional one. We will see below that the
reduced phase space can be described entirely by the
coordinates $r,\alpha, \beta, \xi_1, \xi_2, \xi_3$ or
by the complex coordinates \eq{zeta-s}. The
symplectic structure on the reduced phase space can be derived by the
following Dirac brackets:
\bea
&& \left\{ q_i, q_j \right\}_{DB} =
\left\{ q_i, q_j \right\}_{PB} - \left\{ q_i, f_a \right\}_{PB}\, 
M^{-1}_{ab} 
\, \left\{ f_b, q_j\right\}_{PB}
\nn
&& M_{ab} = \left\{ f_a, f_b \right\}_{PB}
\label{dirac-bracket}
\eea
where $q_i$ refers to one of the six coordinates $r,\alpha, \beta,
\xi_1, \xi_2, \xi_3$ and $f_a$ refers to one of the six constraints
\eq{constraints}. 

The calculation of the Dirac brackets on the supersymmetric subspace
is a straightforward generalization of the half-BPS giant \cite{djm}
and dual-giant \cite{gm} gravitons.  To proceed with the calculations,
let us define the complex coordinates
\be
\label{zeta-s}
\zeta_i = r_i e^{i \xi_i}, \, (r_1,r_2, r_3) \equiv  r (\sin\alpha,
\cos\alpha \sin\beta, \cos\alpha \cos\beta) = r(\mu_1, \mu_2, \mu_3),
\ee
In terms of the coordinates $r_i, \xi_i$ and their conjugate
momenta, the constraints become
\be
f_i \equiv P_{r_i}= 0,\, 
f_{3+i}\equiv P_{\xi_i} - \frac{N }{l^2} r_i^2 =0, \, i=1,2,3
\ee
The non-zero elements of the matrix $M_{ab}$ are given by
\[
\left\{ f_{i}, f_{3+j} \right\}_{PB} = \left \{ P_{r_i}, - 
\frac{N }{l^2} r_j^2
\right\}_{PB} 
= \frac{2 N }{l^2} r_j \delta_{ij}
\]
Thus, for example,
\be
\left\{ \xi_1, r_1 \right\}_{DB} = - \left\{ \xi_1, f_4 \right\}_{PB}
\frac{l^2}{2 N 
  r_1}\left\{f_1, r_1 \right\}_{PB} 
= \frac{l^2}{2 N r_1}
\ee
The Dirac brackets are summarized by
\be
\{\xi_i, r_j^2\}_{DB} = \frac{l^2}{N} \delta_{ij} \implies \left\{
\zeta_i, \bar \zeta_j \right\}_{DB} = -i \frac{l^2}{N}\delta_{ij} 
\ee
\ni\underbar{Symplectic structure}:  
The conclusion of this subsection is that the reduced phase space is
simply ${\mathbf C}^3$ with the symplectic form
\be
\omega = i \frac{N}{l^2}\, d\zeta_i \wedge d\bar \zeta_i
\label{symplectic}
\ee

\subsection{Hamiltonian and charges} 

Since translations along $\xi_i$ are symmetries of the D3-brane action
the momenta $P_{\xi_i}$ are conserved charges. The canonical
hamiltonian, with $P_\alpha = P_\beta = P_r = 0$, becomes
\begin{equation}
H = \frac{1}{l}  \sqrt{ \left(
    \sqrt{\sum_{i=1}^3\frac{P_{\xi_i}^2}{\mu_i^2}}
+ \frac{N r^4}{l^4} \right)^2 +
    \frac{r^2}{l^2} \left( \sqrt{\sum_{i=1}^3
    \frac{P_{\xi_i}^2}{\mu_i^2}} - \frac{Nr^2}{l^2} \right)^2} 
-\frac{Nr^4}{l^5}
\end{equation}
After putting the remaining three constraints from \eq{constraints}
it reduces to
\begin{equation}
H = \frac{Nr^2}{l^3}= \frac N{l^3} \zeta_i \bar \zeta_i
= \frac1 l (P_{\xi_1} + P_{\xi_2} + P_{\xi_3})
\label{SHO}
\end{equation}
which is simply the Hamiltonian of a 3-dimensional simple harmonic
oscillator. 


As expected, the BPS constraint equations automatically satisfy the
equations of motion. The solutions to \eq{constraints} and \eq{SHO}
are
\bea
\label{solspace}
r(t) &=&  r_0,
\nn 
z_k(t) &=& l \,  \mu_k^{(0)} \, e^{i (\xi_k^{(0)} + \frac{t}{l})}
\eea
for $k=1$, $2$ and $3$. The motion is obviously periodic with period
$\Delta t = 2 \pi l$.  Thus we have a
6-dimensional space of solutions parametrized by: $(\mu_k^{(0)}, \,
\xi_k^{(0)}, r_0)$. It is well-known that the space of solutions
of a dynamical system (modulo time-evolution) can be identified with
its phase space.  Folowing Section
\ref{sec:symplectic}, this six-dimensional
space of solutions, viewed as a symplectic space, is ${\mathbb C}^3$.
We will show in the next subsection that the
motion in $S^5$ (second line
of \eq{solspace}) are all in maximal circles in $S^5$ which are related
to the one in $z_1$-plane by $U(3)$ rotations.

\subsection{Interpretation of the solutions as maximal circles}

Let us now try to understand the full solution set of dual-giants
which preserve at least 4 common supersymmetries. A maximal circle on
$S^5$ can be parametrized by 8 parameters. To see this note that every
maximal circle can be obtained by intersecting $S^5$ with an ${\mathbb
R}^2$ passing through the origin in ${\mathbb R}^6$ in which $S^5$ is
embedded. The space of these planes has a dimension of
$\frac{SO(6)}{SO(4) \times SO(2)}$ which is 8. Let us consider
a subspace of maximal  circles
parametrized as $\vec z(\theta)= 
(z_1(\theta),z_2(\theta),z_3(\theta))=  
e^{i \, \theta} \vec z, \, \theta \in (0, 2\pi]$ 
where $z_i$ are
complex coordinates in $C^3$ in which we embed the $S^5$ by $|z_1|^2
+ |z_2|^2 + |z_3|^2 = l^2$. 
Any such circle can be obtained by a
$U(3)$ rotation (defined upto a $U(2)$) 
on a reference circle, for example,
\be
\vec z(\theta) \equiv U e^{i \theta} (1,0,0),
\label{reference}
\ee 
where we can take $U$ to be
\be
U=
\left( \begin{array}{ccc}
e^{i\xi_1}\sin\alpha & - e^{i\xi_1}\cos\alpha & 0\\
e^{i\xi_2}\cos\alpha\sin\beta & e^{i\xi_2}\sin\alpha\sin\beta
&  - e^{i\xi_2}\cos\beta \\
e^{i\xi_3}\cos\alpha\cos\beta & e^{i\xi_3}\sin\alpha\cos\beta
&  - e^{i\xi_3}\sin\beta                    
 \end{array} \right) V.  \;
\label{u3-rotation}
\ee
Here $V \in U(2)$ is an arbitrary matrix that leaves the
column vector $(1,0,0)$ invariant.
Therefore the space of these circles can be identified with
$\frac{U(3)}{U(2)}$. This space has five
real dimensions and gives the parameterization \eq{c-3} of
the unit vector $\vec z$.
We can choose the representation \eq{c-3} for $z_i$.
The time-dependence  of a half-BPS dual-giant on the circle
\eq{reference} is given by putting $\theta = t/l$. Hence
after the $U(3)$-rotation \eq{u3-rotation}, the motion on the generic
maximal circle coincides with the generic $S^5$-motion of 1/8-BPS
dual-giants (see second line of \eq{solspace}).  We conclude that
every dual-giant (of a given size $r_0$) in the full set of
1/8-BPS dual-giant solutions is simply related to every other by a
$U(3)$ rotation.
 
\section{Counting BPS states with $(J_1, J_2,J_3)$}

\subsection{$\frac{1}{8}$-BPS states}

We found in Section \ref{sec:symplectic} that classically the reduced
single-particle phase space is simply ${\mathbb C}^3$, with the
symplectic form \eq{symplectic} and a 3-dimensional simple
harmonic oscillator Hamiltonian, \eq{SHO}.

Since the semiclassical quantization of a simple harmonic oscillator is
exact, quantum mechanically the single-particle Hilbert space is given
by 3-dimensional simple harmonic oscillator eigenstates, viz.
\bea
\ket{n_1, n_2, n_3} &&= \prod_{i=1,2,3}
\frac{(a_i^\dagger)^{n_i}}{\sqrt{n_i !}} \ket 0 
\eea
The hamiltonian and the charges are given by
\bea
P_{\xi_i} && \equiv J_i =  n_i, \, i=1,2,3
\nn
l H &&= n_1 + n_2 + n_3 
\label{charges}
\eea 
Here the classical phase space variables $\zeta_i$, $\bar \zeta_j$ are
quantized as $(l/\sqrt{N})~ \zeta_i \to a_i$, $(l/\sqrt{N}) ~ \bar
\zeta_j$ $\to a^\dagger_j$. The operator ordering of the charges has
been obtained by generalizing from the half-BPS case \cite{dms}; in
the Hamiltonian we have dropped the zero point energy which is not
important for our purposes. We note that the conserved momenta
$(P_{\xi_1},P_{\xi_2},P_{\xi_3}) $ by construction correspond to the
three angular momenta $(J_1, J_2, J_3)$ of $SO(6)$.  In what follows
we will use the notation $J_i$ in stead of $P_{\xi_i}$.

As we have argued earlier, putting any number of particles in this
reduced Hilbert space is consistent with 1/8 supersymmetry.  Since
they are BPS objects (with respect to each other) the total energy is
given by the sum of the individual energies (like in the case of
half-BPS states). Further we can have more than one dual-giants with
exactly the same quantum number and so they can be treated as bosonic
objects. Since each dual giant being an $S^3$ inside $AdS_5$ acts as a
domain wall and therefore considerations of \cite{bs, nvs} apply. This
restricts the maximum number of dual-giants to $N$. The total angular
momenta should also be given by the sum of those for the individual
dual-giants:
\begin{equation}
J_i = \sum_{k=1}^N J_i^{(k)} ~~~ \hbox{for} ~~~ i =1, ~ 2,
~3. 
\end{equation}
From \eq{charges} it is easy to see that the partition function for
the dual-giant graviton system, consistent with 1/8 supersymmetries,
is given by that for $N$ bosons in a 3-dimensional simple harmonic
oscillator. Here we can identify these bosons with dual-giant
gravitons. By including the $J_i =0$ state we may simply count all
states with a total of $N$ bosons with some of them sitting at the
zero energy state. This takes care of the configurations with less
than $N$ dual-giants. The grand canonical partition function is,
therefore, 
\be {\cal Z}( \zeta, q_1, q_2, q_3 ) \equiv Tr \exp[- \mu N
- \beta_i J_i] = \prod_{n_1=0}^\infty \prod_{n_2=0}^\infty
\prod_{n_3=0}^\infty (1 - \zeta q_1^{n_1} q_2^{n_2} q_3^{n_3})^{-1}
\label{z-1/8}
\ee
The chemical potentials $q_i \equiv e^{-\beta_i}$ are conjugate to
the charges $J_i$ and  the `fugacity' $\zeta = e^{ -\mu}$ 
is conjugate to number $N$ of dual-giants:
\begin{equation}
{\cal Z} (\zeta, q_1, q_2, q_3) = \sum_{N=0}^\infty \zeta^N Z_N(q_1,
  q_2, q_3) 
\end{equation}
with 
\begin{equation}
Z_N(q_1, q_2, q_3) = \sum_{J_1, J_2, J_3 = 0}^\infty \Omega_N (J_1,
J_2, J_3) q_1^{J_1} q_2^{J_2} q_3^{J_3}.
\end{equation}
Some special configurations however have enhanced supersymmetries to
either 1/4 or half-BPS states and so we may choose to subtract them
out.  For this a useful way to think about this counting is the
following. A generic state can be specified by $N$ identical bosonic
particles sitting on the 3-dimensional lattice with each lattice point
coordinates $n_1,n_2,n_3$ which take integer values.The half-BPS
states again are the configurations of points on this 3-dimensional
lattice where all $N$ points fall on a single line going through the
origin. The 1/4-BPS states are those for which all $N$ points fall on
a single 2-plane going through the origin again. We will discuss
counting of 1/4-BPS states separately in the next subsection.

\subsection{$\frac{1}{4}$-BPS states}

As stated above, dual-giant configurations for which all $N$
dual-giants have the same specific $\alpha$ and $\beta$ preserve 1/4
of the supersymmetries (eight supersymmetries). In terms of the bosons
on a 3-dimensional lattice these are the configurations for which all
bosons lie on the same 2-plane passing through the origin. The choice
of which eight supersymmetries we want to preserve fixes the values of
$\alpha$ and $\beta$ (and therefore fixes a plane in the 3-d
lattice). For the purpose of counting, we may choose this plane to
be the $\mu_3 =0$ plane ($\beta=\pi/2$), which in turn fixes
$P_{\xi_3}=0$. Classically,
\begin{equation}
P_{\xi_1} = \frac{Nr^2}{l^2} \sin^2\alpha, ~~~ P_{\xi_2} =
\frac{Nr^2}{l^2} \cos^2\alpha.
\end{equation}
Repeating the same steps as in Section \ref{sec:symplectic}, 
we can now find a four-dimensional phase space with the
symplectic structure of ${\mathbf C}^2$, expressed
by the following Dirac brackets
\begin{equation}
\left\{\xi_1, r^2 \sin^2\alpha \right\}_{DB} = \frac{l^2}{2N}, 
~~\left\{ \xi_2, r^2
\cos^2\alpha \right\}_{DB} = \frac{l^2}{2N}.
\end{equation}
As before, for a fixed $\alpha$ they preserve 16 supersymmetries. But
if we put together dual-giants with various values of $\alpha$ that
configuration preserves only 8 supersymmetries.

Quantization proceeds in a manner similar to the 1/8-BPS
case. The conserved momenta $(P_{\xi_1}, P_{\xi_2})$ are
identified as the quantized angular momenta $(J_1, J_2)$.
Counting  the 1/4-BPS states goes as follows. To
make a 1/4-BPS state we need to have at least two dual-giants in that
configuration with different values of $\alpha$. The specific values
of $\alpha$ should be such that we get integer values for $J_1$
and $J_2$. The other constraint is that one can have a total of
no more than $N$ dual-giant gravitons.

So the 1/4-BPS states can be counted by considering the number of
ways, $\Omega_N(J_1, J_2)$, in which one can distribute $N$ identical
bosons on a 2-dimensional lattice of integers $(n_1, n_2)$, such that
the sum of their individual $n_1$'s is $J_1$ and the sum of the
individual $n_2$'s is $J_2$.  A partition function which generates
this number is
\bea
{\cal Z}(\zeta,q_1,q_2) 
&&= \prod_{n_1,n_2=1}^\infty (1- \zeta \, q_1^{n_1}
q_2^{n_2})^{-1}
= \sum_{N=0}^\infty Z_N (q_1, q_2) \zeta^N  
\label{z-1/4}
\eea
%
%
The above counting includes states which preserve at least 1/4 of the
supersymetries.  To exclude the special states which preserve 1/2 of
the supersymmetries one needs to exclude configurations in which all
particles lie on a straight line passing through the origin.  Each
line passing through the origin in a 2-dimensional lattice can be
uniquely specified by a pair of relatively prime integers, say
$(n_1,n_2)=(r,s)$. This line $(r,s)$ corresponds to the points
$(n_1,n_2)=k\, (r,s), \, k=1,...,\infty$. Combining the contribution
of all such lines, the part of \eq{z-1/4} which corresponds to
1/2-BPS states, is given by
\begin{equation}
Z^{(N)}_{\rm 1/2-BPS} (q_1, q_2) = 
\prod_{r, s \ge 0 \atop {{\rm gcd}(r,s)=1 \atop (r,s) \ne
(0,0)}} \prod_{k=1}^N \frac{1}{1-{(q_1^r q_2^s)}^k}
\label{z-1/2}
\end{equation}
So to eliminate the contribution from the states with enhanced
supersymmetries one should simply subtract \eq{z-1/2} from $Z_N(q_1,
q_2)$ defined in \eq{z-1/4}.

\subsection{Comparison with gauge theory answers}

Single trace 1/4-BPS operators of ${\cal N} = 4$ SYM with $SU(N)$
gauge group have been constructed in the literature
\cite{ryzhov,dhhr,bks}. These states belong to the $[p,q,p]$
representation of $SU(4)$ which has $(J_1, J_2) = (p+q, p)$.  More
generally, an index for ${\cal N}=4$ Yang-Mills theories has recently
been calculated \cite{kmmr} (see also \cite{cr}) which counts 1/8-,
1/4- and 1/2-BPS states of the kind we discussed above. Our results
\eq{z-1/8} and \eq{z-1/4} agree with their result. In case of the
1/8-BPS states, the dual-giant graviton states we constructed above
are to be identified with gauge-invariant operators which do not
involve the fermionic fields.

\section{1/8-BPS states with $(S_1, S_2, J_1)$}

In this section we consider a different problem, namely that of
configurations of multiple giant gravitons which carry non-zero $(S_1,
S_2, J_1)$ charges and preserve at least 4 supersymmetries. Similar
configurations have been considered in the literature before in
\cite{adkss, cs}. We work in the coordinates that we used in the
earlier sections and consider configurations of  D3-branes of the
type:
\begin{eqnarray}
&& t = \tau, ~~, r= r(\tau), ~~ \theta = \theta(\tau), ~~ 
\phi_1 = \phi_1(\tau), ~~ \phi_2 = \phi_2(\tau), \cr
&& \alpha= \alpha(\tau), ~~\beta = \sigma_1, ~~ \xi_1 = 
\xi_1 (\tau), ~~ \xi_2 = \sigma_2, ~~
\xi_3 = \sigma_3
\end{eqnarray}
The pull-back of the RR 4-form is $C^{(4)}_{\tau \sigma_1 \sigma_2
  \sigma_3} = - l^4 \, \cos^4 \alpha \, \sin \sigma_1 \, \cos
  \sigma_2$. Then the D3-brane Lagrangian becomes
\begin{equation}
L = - \frac{N \, \cos^3 \alpha}{l}\left[ \Delta^{1/2} - l \,
  \cos\alpha \, \dot \xi_1 \right]
\end{equation}
where 
\begin{equation}
\Delta = V(r) - \frac{{\dot r}^2}{V(r)} - r^2 \, ( \dot \theta^2 +
\cos^2\theta \, \dot \phi_1^2 + \sin^2\theta \, \dot \phi_2^2) 
- l^2 (\dot\alpha^2 + \sin^2\alpha \, \dot \xi_1^2)
\end{equation}
Notice the change of sign for the Chern-Simons term. The reason for
this is we have chosen here an anti-D3-brane rather than a D3-brane so
as to find a solution with positive sign of $\dot \xi_1$. It is
easy to see that the following configurations satisfy the equations of
motion
\begin{equation}
\dot\alpha= \dot r = 
\dot \theta = 0, ~~ |\dot \phi_1| = |\dot \phi_2| = \dot
\xi_1 = 1/l
\label{ggsolns}
\end{equation}
In fact these configurations saturate a Bogomolny bound and the
Hamiltonian reads (for positive values of $l \, \dot
\phi_1$, $l \, \dot \phi_2$)
\begin{equation}
H = \frac{1}{l} (P_{\phi_1} + P_{\phi_2} + P_{\xi_1})
\end{equation}
where 
\be
\label{gg-momenta}
P_{\xi_1} = N \, \cos^2 \alpha, P_{\phi_1} =
(r^2/l^2) \, P_{\xi_1} \, \cos^2 \theta, P_{\phi_2} = (r^2/l^2)\,
P_{\xi_1} \, \sin^2\theta. 
\ee
We will show in the next
subsection that the configurations \eq{ggsolns} 
share at least 4 supersymmetries for arbitrary values
of $r$, $\theta$ and $\alpha$.
In terms of the canonical variables, the supersymmetry
conditions \eq{ggsolns} imply the following constraints 
\begin{eqnarray}
&& P_r = 0, ~~ P_\theta = 0, ~~ P_\alpha = 0, ~~ P_{\phi_1} -
  \frac{r^2}{l^2} \, N \, \cos^2\alpha \, \cos^2\theta = 
  0, \cr
&& P_{\phi_2} - \frac{r^2}{l^2} \, N \, \cos^2\alpha \,
  \sin^2\theta = 0, ~~ P_{\xi_1} - N \, \cos^2\alpha= 0.
\label{gg-constraints}
\end{eqnarray}
As for the case of the dual giant gravitons earlier, we can define 
$$(r_1, r_2, r_3) = (r \, \cos\alpha \, \cos\theta, r\, \cos\alpha \,
\sin\theta, l \, \cos\alpha)$$
and 
$$\zeta_1 = r \, \cos\alpha \, \cos\theta \, e^{i \phi_1}, ~ \zeta_2 =
r \, \cos\alpha \, \sin\theta \, e^{i\phi_2}, ~ \zeta_3 = l \,
\cos\alpha \, e^{i \, \xi_1}$$
Because of the six constraints \eq{gg-constraints} the 12-dimensional
phase space of the $\zeta_i, \bar \zeta_i$ and their conjugate momenta
gets reduced to a six-dimensional phase space. The problem of finding
the Dirac brackets in the reduced phase space is similar to the
previous case and we get
\begin{equation}
\left\{\zeta_i, \bar\zeta_j\right\}_{DB} 
= -i \, \frac{l^2}{N} \, \delta_{ij}
\label{gg-dirac}
\ee
which gives the same symplectic form, Eq. \eq{symplectic},
as before.  

The main difference of this solution space from the earlier one is
that $|\zeta_3| \le l$. So the phase space is really ${\mathbf C^2}
\times D$ with the symplectic form \eq{symplectic} inherited from
${\mathbf C^3}$. Note that the boundary is a null curve of the
symplectic form which is as it should be.

In terms of the new variables, \eq{gg-momenta} reads
\begin{equation}
P_{\phi_1} = \frac{N}{l^2} \, | \zeta_1|^2, ~~ P_{\phi_2} =
\frac{N}{l^2} \, |\zeta_2|^2, ~~ P_{\xi_1} = \frac{N}{l^2} \,
|\zeta_3|^2 
\end{equation}
and the Hamiltonian written in these coordinates becomes
\begin{equation}
l \, H =  \frac{N}{l^2} \sum_{i=1}^3 \zeta_i \bar \zeta_i
\label{gg-ham}
\end{equation}
Eqs. \eq{gg-ham}, \eq{gg-dirac} imply that  
the system is again a 3-dimensional simple harmonic oscillator
(the implication of bounded $|\zeta_3|$ for quantization
will be discussed shortly).

\subsection{Supersymmetries of spinning giants}

The world-volume gamma matrices are
\begin{eqnarray}
&& \gamma_\tau = V^{1/2} \, \Gamma_0 + \frac{\dot r}{V^{1/2}} \,
  \Gamma_1 + r\, \dot \theta \, \Gamma_2 + r \, (\dot \phi_1 \,
  \Gamma_3 \, \cos\theta + \dot \phi_2 \, \Gamma_4 \, \sin\theta) + l
  \, (\dot \alpha \, \Gamma_5 + \sin\alpha \, \dot \xi_1 \, \Gamma_7),
  \cr 
&& \gamma_{\sigma_1} = l \, \cos\alpha \, \Gamma_6, ~~
  \gamma_{\sigma_2} = l \, \mu_2 \, \dot \xi_2 \, \Gamma_8, ~~
\gamma_{\sigma_3} = l \, \mu_3 \, \dot \xi_3 \, \Gamma_9.
\end{eqnarray}
The kappa projection equations for an anti-D3-brane are
\begin{eqnarray}
&& [ V^{1/2} \, \Gamma_0 + \frac{\dot r}{V^{1/2}} \,
  \Gamma_1 + r\, \dot \theta \, \Gamma_2 + r \, (\dot \phi_1 \,
  \Gamma_3 \, \cos\theta + \dot \phi_2 \, \Gamma_4 \, \sin\theta) \cr
&& ~~~~~~~~~~ \, + l
  \, (\dot \alpha \, \Gamma_5 + \sin\alpha \, \dot \xi_1 \, \Gamma_7)
  - i \, \Delta^{1/2} \, \Gamma_{689}] \epsilon = 0
\end{eqnarray}
On the solutions $\dot r = 0$, $\dot \theta = 0$, $\dot \alpha = 0$,
$\dot \phi_1 = \dot \phi_2 = \dot \xi_1 = 1/l$ this reduces to 
\begin{equation}
[ V^{1/2} \, \Gamma_0 + \frac{r}{l} ( \Gamma_3 \, \cos\theta +
  \Gamma_4 \, \sin\theta) + \Gamma_7 \, \sin\alpha - i \Gamma_{689} \,
  \cos\alpha ] \, \epsilon = 0.
\end{equation}
To simplify this equation we use the following identity
\begin{eqnarray}
(\sin\alpha \, \Gamma_7 - i 
\, \cos\alpha \, \Gamma_{689})M = -i \, M
\, \Gamma_{57} \tilde \gamma
\end{eqnarray}
Then the killing spinor equation can be rewritten as
\begin{eqnarray}
&& M[V \, \Gamma_0 + \frac{r}{l} V^{1/2} \, \cos\theta \, e^{-\phi_1 \,
    \Gamma_{13} - i \frac{t}{l} \Gamma_0 \, \gamma} (\Gamma_3 + i \,
    \Gamma_{01} \, \gamma) \cr
&&  + \frac{r}{l} V^{1/2} \, \sin\theta \,
    e^{-\phi_1 \, \Gamma_{24} -i \frac{t}{l} \Gamma_0 \, \gamma}
    (\Gamma_4 + i \, \Gamma_{02} \, \gamma) - i \frac{r^2}{l^2}
    (\cos^2\theta \, \Gamma_{13} + \sin^2\theta \, \Gamma_{24}) \gamma
    \cr
&&  - i \, \frac{r^2}{l^2} \cos\theta \, \sin\theta \, (\Gamma_{23} +
    \Gamma_{14}) \, \gamma \, e^{-\phi_1 \, \Gamma_{13} - \phi_2 \,
    \Gamma_{24}} - i \, \Gamma_{57} \, \tilde \gamma ] \, \epsilon_0 = 0.
\end{eqnarray}
Since we are interested in having configurations of multiple giant
gravitons with generically different values of $(r, \theta, \alpha)$
we may impose the projections
\begin{equation}
\label{sgprojections}
(\Gamma_0 - i \, \Gamma_{13} \gamma) \, \epsilon_0 = 0, ~~ (\Gamma_0 -
  i \, \Gamma_{24} \gamma) \, \epsilon_0 = 0, ~~ (\Gamma_0 - i\,
  \Gamma_{57} \, \tilde \gamma) \, \epsilon_0 = 0.
\end{equation}
with which the killing spinor equation is satisfied. So the solutions
we have found share at least 4 supersymmetries.

Using methods similar to the ones used in section (2.3) one may show
that the full set of solutions given the projections in
(\ref{sgprojections}) is the one considered above.

Further one can interpret the solutions as all those obtained by
$U(1,2)$ rotations acting on the homomorphic coordinates $(\Phi_0,
\Phi_1, \Phi_2)$ of ${\mathbb C}^{1,2}$ (in which $AdS_5$ is embedded
as $-|\Phi_0|^2 + |\Phi_1|^2 + |\Phi_2|^2 = - l^2$) on the original
half-BPS giant graviton. 

\subsection{Quantization}

The phase space is ${\mathbf C^2} \times D$ where $D$ represents
the Disc $|\zeta_3| \le 1$. 

The $D$ part of the phase space (with the symplectic form and the
Hamiltonian) is identical to that of the half-BPS giant gravitons
\cite{dms}. The semiclassical quantization corresponds to
a fuzzy disc and the exact single-particle quantum mechanics is described by
$H_{1,N}$ which is an N-dimensional Hilbert space
consisting of the first N levels of an simple harmonic oscillator \cite{dms}.
\be
{\cal H}^{1,N} = \hbox{Span} \{ \ket n, n=1,2,...,N \}
\ee
We have omitted the $n=0$ state from the spectrum of giant gravitons
since the minimum angular momentum of a half-BPS giant graviton is
unity. 
The single-particle quantum mechanics corresponding to
${\mathbf C^2}$ is given by the states of a 2D simple harmonic oscillator of
arbitrarily high quantum numbers:
\be
{\cal H}' = \hbox{Span} \{ \ket{l, m}, l,m=0,1,...,\infty \}
\ee
The full single-particle Hilbert space of the giant
gravitons is given by 
${\cal H}_1={\cal H}^{1,N} \times {\cal H}'$.

The three angular momenta \eq{gg-momenta} are identified now
as the three charges $(S_1, S_2, J_1)$ where the first two
correspond to angular momenta in AdS$_5$ and $J_1$ correspond
to angular momentum in $S^5$.
\begin{equation}
\label{ssjptfn}
Z(q_1,q_2,q_3) =  \prod_{l,m=0}^\infty \prod_{n=1}^N 
\frac{1}{1-q_1^l
  \, q_2^m \, q_3^n} = \sum_{S_1, S_2, J_1 = 0}^\infty \Omega_N(S_1,
  S_2, J_1) \, q_1^{S_1} \, q_2^{S_2} \, q_3^{J_1}
\end{equation}
Similar to the situation in the 1/8-BPS states with $(J_1, J_2, J_3)$
one has configurations in this partition function which have more than
4 supersymmetries. For example whenever there is just one
giant-graviton making up a state, since it can be obtained by an
isometry of $AdS_5$ acting on the standard half-BPS giant graviton it
is expected to preserve 16 supercharges \cite{cs} (see also
\cite{adkss}). Further whenever all the giants in a given state have
the same value of $r$ that state is expected to preserve 8
supercharges. Again if one wishes one can systematically exclude the
contribution of these states from (\ref{ssjptfn}) to get the
degeneracies of exactly 1/8 or 1/4 supersymmetric states.

\section{Conclusions}

In this paper we considered the counting problem of quantum states in
the type IIB string theory on $AdS_5 \times S^5$ background that
preserve at least 4 supercharges (1/8-BPS). Two types of states have been
considered: 
\begin{itemize}
\item those with non-zero $(J_1, J_2, J_3)$ with $E = J_1 +
J_2 + J_3$ where $E$ is their energy conjugate to the time coordinate
in global coordinates and $J_i$ are the three independent angular
momenta on $S^5$,

\item those with non-zero $(S_1, S_2, J_1)$ with
$E= S_1 + S_2 + J_1$ where $S_i$ are the two independent angular
momenta on $S^3 \subset AdS_5$. 
\end{itemize}
For the first set we considered $N$-particle states of dual-giant
gravitons rotating along arbitrary maximal circles of $S^5$ that share
at-least four supersymmetries. The result can be expressed quite simply
in terms of the degeneracy of states of an $N$-boson system in a
three-dimensional harmonic oscillator potential. 

For the second set of 1/8-BPS states we considered configurations
containing an arbitrary numbers of giant gravitons having angular
momenta in the $AdS_5$ directions which share at least 4
supersymmetries. The result for the counting problem for this set of
states can again be mapped onto that of a 3-dimensional harmonic
oscillator, but this time with an arbitrary number of bosons
representing the giant gravitons and with one of the three quantum
numbers of the 3-dimensional harmonic oscillator bounded above by $N$.

The first type of 1/8-BPS states can also be given by the
classification of giant gravitons of \cite{mikhailov}. Recently
\cite{bglm} have counted these 1/8-th BPS giant graviton
configurations by quantizing Mikhailov's solutions. The counting
problem in terms of giant gravitons appears to be significantly more
complicated since the giant gravitons wrap different 3-surfaces within
the $S^5$ which can have complicated intersections and quantization
involves quantizing the space of these 3-surfaces. The dual giant
gravitons, however, have a much simpler description since their
world-volume, at any given time, is a three-sphere of a given size
inside $AdS_5$ and their motion is that of point particles on the
$S^5$. Remarkably, both these descriptions give the same counting of
1/8-th BPS states. This points to some duality between the giant
graviton and the dual giant graviton descriptions.

It is pertinent to recall at this point \cite{nvs, dms} that in case
of half-BPS states there is an explicit duality between giant graviton
states and dual giant graviton states. If we specify a multi-giant
graviton state by $(r_1, r_2, \cdots, r_N)$ where $r_i$ denotes the
number of giant gravitons with angular momentum $J_1 = i$ and a
multi-dual-giant graviton state by $(s_1, s_2, \cdots, s_N)$ where
$s_i$ is the angular momentum of the i$^{th}$ (counting
from the dual-giant with largest $J_1$) then the duality map is given
by
\be
s_i = \sum_{k=i} r_k
\label{half-bps-duality}
\ee
Thus, the number of dual-giants becomes the number of single-particle
energy levels for giants. Also, the occupied energy levels ($s_i$) of
dual-giants get related \eq{half-bps-duality} to occupation numbers
($r_k$) of individual levels in the giant graviton system. It will be
very interesting to explore if this type of a duality exists for the
1/8-BPS states with $(J_1, J_2, J_3)$ quantum numbers considered here
and in \cite{mikhailov, bglm}. Further in the half-BPS sector the
duality between giants and dual-giants follows form a unified
description of the system \cite{db1, llm} in terms of $N$
fermions in a harmonic oscillator potential. It will be interesting to
see if such a unified picture can be given for lower supersymmetric
cases too. It is possible that such a description arises in the
solution of matrix models proposed in \cite{db2} to capture the
physics of 1/8-BPS states.

Similar to the description of giant gravitons in \cite{mikhailov} one
can describe a dual-giant graviton as a three surface obtained by
intersection of $\Phi_0 = \varphi_0^{(0)}$ and $Z_1 = \zeta_1^{(0)}$,
$Z_2 = Z_3 = 0$ with ${\mathbb C}^{1,2} \times {\mathbb C}^3$ and
$-|\Phi_0|^2 + |\Phi_1|^2 + |\Phi_2|^2 = -l^2$ and $\sum_{i=1}^3
|Z_i|^2 = l^2$ evolving in time as $\Phi_0 \rightarrow \Phi_0 \,
e^{i\frac{t}{l}}$ and $Z_1 \rightarrow Z_1 \, e^{i \frac{t}{l}}$ where
$(\Phi_0, \Phi_1, \Phi_2)$ are holomorphic coordinates on ${\mathbb
C}^{1,2}$ and $(Z_1, Z_2, Z_3)$ are holomorphic coordinates on
${\mathbb C}^3$ and $\varphi_0^{(0)}, ~ \zeta_1^{(0)}$ are arbitrary
complex numbers. This can be generalized to `wobbling dual-giants'
\cite{ms}
\begin{equation}
g(\Phi_0, \Phi_1, \Phi_2) = 0, ~~ Z_1 = \zeta_1^{(0)}, ~~Z_2 = Z_3 = 0 
\end{equation}
with the time evolution $\Phi_i \rightarrow \Phi_i \, e^{i \,
\frac{t}{l}}$ and $Z_1 \rightarrow Z_1 \, e^{i\, \frac{t}{l}}$. These
states carry non-zero $(S_1, S_2, J_1)$ generically and can be shown
to preserve at least 1/8 of the supersymmetries of the background
\cite{ms}. One should be able to quantize the space of these `wobbling
dual-giants' and count the 1/8-BPS states using the methods of
\cite{bglm}. It will be interesting to see if there exists a duality
between our giant-graviton configurations and these wobbling
dual-giants our counting results gives a prediction for the dual-giant
counting.

In both types of 1/8-BPS states we considered here they preserve
$SO(4)$ symmetry (coming from isometries of $S^3 \subset AdS_5$ for
the states with $(J_1, J_2, J_3)$ charges and from $S^3 \subset S^5$
for those with $(S_1, S_2, J_1)$ charges). So different giants in a
given state form concentric three-spheres and never intersect (and
whenever they do they actually coincide). Usually D-branes which
intersect are expected to split and rejoin and therefore the
degeneracies of such states can change. But in our case since they do
not intersect we suggest that their degeneracies should not receive
any quantum corrections. It will be interesting to see if one can give
a description of the full set of 1/8-BPS states with given charges in
our language by turning on some bosonic or fermionic zero modes on the
world-volume of the probe branes that break the $SO(4)$ invariance but
not supersymmetry.

It is of interest to find an exact orthonormal basis of states in the
dual ${\cal N} = 4$ $U(N)$ SYM for the 1/8-BPS states considered
here. For half-BPS states an orthonormal basis of operators in SYM was
provided in \cite{cjr}. See \cite{cs} for some comments on the dual
operators for the states with non-zero $(S_1, S_2, J_1)$.

\section*{Acknowledgements}
We would like to thank Shiraz Minwalla and Rob Myers for helpful
discussions. We thank Shiraz Minwalla for sharing with us the results in
\cite{bglm} prior to publication. NVS would like to thank TIFR,
Mumbai for hospitality during the final stages of this work.

\end{document}